\documentclass{article}
\usepackage{graphicx}
\usepackage[font=small,skip=0pt,justification=raggedright]{caption}
\usepackage{amssymb}
\usepackage{mathrsfs}
\usepackage{verbatim}
\usepackage{dutchcal}
\usepackage{comment}
\usepackage{xcolor}

\usepackage{url}
\usepackage{lineno}
\usepackage{url}
\usepackage{subcaption} 

\usepackage{epsfig}
\usepackage{amsmath}
\usepackage{amsfonts}\tolerance=10000
\date{\today}

 \def\a{\alpha}

\newcommand{\insertplot}[5]{\begin{figure}
 \hfill\hbox to 0.05in{\vbox to #5in{\vfill
 \inputplot{#1}{#4}{#5}}\hfill}
 \hfill\vspace{-.1in}
 \caption{#2}\label{#3}
 \end{figure}}
 \newcommand{\inputplot}[3]{
 \special{ps: plotfile #1}
}
\newcounter{fig}   
\newcommand{\vphi}{\varphi}

\newcommand{\beq}{\begin{equation}}
\newcommand{\eeq}{\end{equation}}
\newcommand{\beqs}{\begin{eqnarray}}
\newcommand{\eeqs}{\end{eqnarray}}

\numberwithin{equation}{section}
\newcommand{\be}{\begin{equation}}
\newcommand{\ee}{\end{equation}}
\newcommand{\bea}{\begin{eqnarray}}
\newcommand{\eea}{\end{eqnarray}}

\usepackage{graphicx}

\begin{document}

\title{
  Reissner-Nordstr\"om dyonic black holes 
  with gauged scalar hair
}

\author{
{\large Carlos Herdeiro}$^{1}$,
{\large Eugen Radu}$^{1}$
and
{\large Yakov Shnir}$^{2,3,4}$,
\\
\\
$^{1 }${\small Departamento de Matem\'atica da Universidade de Aveiro and }
\\ {\small  Centre for Research and Development  in Mathematics and Applications (CIDMA),}
\\ {\small    Campus de Santiago, 3810-183 Aveiro, Portugal}
\\
$^{2}${\small BLTP, JINR, Dubna 141980, Moscow Region, Russia}
\\
$^{3}${\small Institute of Physics, Carl von Ossietzky University Oldenburg,
Germany Oldenburg D-26111, Germany}
\\
$^{4}${\small Hanse-Wissenschaftskolleg, Lehmkuhlenbusch 4, 27733 Delmenhorst, Germany}
}

\date{June 2024}

\maketitle
 
\begin{abstract}
For gauged scalar fields minimally coupled to Einstein-Maxwell theory,  the 
Mayo-Bekenstein no-hair theorem 
 can  be circumvented when including appropriate scalar self-interactions,
allowing static, 
electrically charged
black holes
to be endowed with (Abelian) gauged scalar hair.
Here we show these spherically symmetric solutions can be extended to include
a magnetic charge in  a model
with scalar multiplets.
The resulting dyonic configurations share most of the properties of 
the electrically charged solutions,
in particular satisfying the same
{\it resonance} condition,
 with 
 the existence of a mass gap with respect to the bald
 Reissner-Nordstr\"om dyonic black holes.
A distinctive feature, however, is that no solitonic limit exists for a non-zero
 magnetic charge.
 
\end{abstract}


 \section{Introduction }
  \label{Introduction}

Even if magnetic monopoles are not realized in the physical world, 
there is surely much to be gained,
in both physics and mathematics,
by studying the theory
of monopoles  - for reviews, see, 
\cite{shnir2006magnetic,Manton:2004tk}.
In the simplest setting, 
these hypothetical particles naturally occur 
as sources in the electric-magnetic symmetric extension of
Maxwell's equations.
Moreover, 
their existence leads to far reaching consequences
within the framework of quantum mechanics,
since, as found by Dirac \cite{Dirac:1931kp},
it would explain electric 
charge quantization, 
with the simple relation
(where we consider natural units with $\hbar=c=1$):
 \begin{eqnarray}
    \label{Dirac}
q Q_m=\pm \frac{N}{2},~~~N=0,1,2\dots~,
\end{eqnarray}
  $Q_m$ being the magnetic charge and 
$q$  the gauge coupling constant, $i.e.$ the electric
charge of field quanta.

The theory of Abelian monopoles is somewhat obscured by
the unphysical string-like singularity\footnote{This singularity is removed
in the Wu-Yang formulation of theory \cite{Wu:1975es}.} 
of the Dirac potential
\cite{Dirac:1931kp}, as well as a rather cumbersome modification of the Lagrangian formulation of theory.
Another undesirable feature is  
the existence of 
a central singularity\footnote{ 
The situation is different in models with non-Abelian fields.
As found independently by 't Hooft~\cite{tHooft:1974kcl} and Polyakov~\cite{Polyakov:1974ek},
globally regular, particle-like
solutions with a magnetic charge may exist in
$SU(2)$-Yang-Mills-Higgs theory. 
Magnetic monopoles may exist also in the Standard Models of particle
physics, although they possess a central singularity and an infinite mass
\cite{Cho:1996qd}.
},
which, however, can be hidden beyond an event horizon 
when including General Relativity effects.
In the absence of rotation,
the uniqueness theorems
\cite{Chrusciel:2012jk}
show that  the resulting configuration 
corresponds to a Reissner-Nordstr\"om  (RN)
(magnetic) black hole (BH).

The interaction between a magnetic monopole and a ($U(1)$-gauged) scalar field is a well known problem, see $e.g.$   
\cite{boulware1976scattering}.
Nevertheless, 
most of the studies  
have focused on
the case of a flat spacetime background.
 In particular, 
 to the best of our knowledge,
 the possible coexistence
 of  $U(1)$-gauged scalar fields with a  
RN magnetic BH
has  not been addressed 
 in the literature.

In this context, a central result is the Mayo-Bekenstein no-hair theorem
\cite{Mayo:1996mv}, showing the absence of gauged scalar hair for \textit{electric} RN and scalar fields \textit{without} self-interactions. A few years ago, however, it was pointed out that
the  Mayo-Bekenstein theorem
$can$ be circumvented
and a static, spherically symmetric,
electrically charged RN BH
can be endowed with (Abelian) gauged scalar hair in the presence of \textit{appropriate scalar self-interactions}~\cite{Hong:2019mcj,Herdeiro:2020xmb,Hong:2020miv}.
This result came somehow as a surprise, since
for a model with a global $U(1)$ invariance
($i.e.$ an ungauged complex scalar field)
the only spherically symmetric solution\footnote{Here, a number of physical
conditions are assumed, but self-interactions are allowed - see
the review work \cite{Herdeiro:2015waa}.}
is the
Schwarzschild BH \cite{Pena:1997cy}.

The hairy BH  configurations
in 
\cite{Hong:2019mcj,Herdeiro:2020xmb,Hong:2020miv}
 require the scalar field to be self-interacting, include a mass term and vanish at infinity,
which excludes the case
of a Higgs field.  
Their existence relies on the \textit{resonance condition}:
\begin{eqnarray}
    \label{cond}
    \omega=q \Phi ~,
\end{eqnarray}
where
$\omega>0$ is the scalar field frequency and
$\Phi$
is the electrostatic chemical potential.
The hairy BHs possess a non-trivial horizonless (i.e. solitonic) limit, 
reducing to spherical,
 gauged boson stars 
\cite{Jetzer:1989av,Jetzer:1992tog,Pugliese:2013gsa}.\footnote{A similar feature occurs for the two-component scalar Einstein-Maxwell-Friedberg-Lee-Sirlin model with a symmetry breaking potential
\cite{Kunz:2023qfg}.} 

Electric-magnetic duality
is a symmetry of Einstein-Maxwell theory, making electric and magnetically charged BHs belong to the same symmetry orbit~\cite{Hawking:1995ap,Herdeiro:2020iyi}. Since electric magnetic-duality is not a symmetry of the Einstein-Maxwell-scalar model allowing for the aforementioned hairy BHs,  one may ask how the results above
are affected by the presence of
a magnetic charge. 
In particular: (i) do such hairy 
solutions
exist in the 
absence of an electric charge?
 (ii) What is the generalization
 of  the  resonance condition
(\ref{cond})
  for $Q_m \neq 0$?
  
\medskip

This work is a step
to address these questions. We shall restrict our study to the case of
a static, spherically symmetric spacetime.
We first prove the absence 
of linear scalar clouds ($i.e.$ bound states)
on a dyonic RN background,
in particular for a purely magnetic one.
This does not exclude, however, the 
existence of non-linear solutions, which are the only clouds allowed in the $Q_m =0$
limit
\cite{Hong:2019mcj,Herdeiro:2020xmb,Hong:2020miv}.
Since a
 configuration with a single
 gauged scalar field 
cannot be spherically symmetric
in the presence of a  magnetic charge,
we next consider 
 a model with
{\it scalar multiplets}.
There we employ  a specific ansatz which factorizes the
angular dependence,
such that the total energy-momentum tensor is
 consistent with a spherically symmetric geometry.
The solutions presented - all of them numerical - are found 
for the simplest case
of a model with two  scalars
 and a unit magnetic charge monopole.

 Using this framework, we 
 prove 
that
the  RN  dyonic BH
($i.e.$ with both electric and magnetic charges)
possesses  generalizations with scalar hair.
The existence of hairy solutions is still anchored in the
 resonance condition (\ref{cond}),
which is not affected by the  presence
 of a magnetic charge.
Moreover,
as in the purely electric case,
the existence of hair
requires the scalar field to be massive
\textit{and} self-interacting; no such solutions exist for  massive but free scalar fields.
As a new feature, we find that 
a smooth horizonless solitonic limit of the hairy BHs
ceases to exist in
the presence of a magnetic charge.
 
\section{The action and field equations }

We consider the $D=3+1$ Einstein's gravity minimally coupled to a Maxwell field  and
a set of $n$
massive, complex, gauged scalar fields $\Psi^{(k)}$.
The Einstein-Maxwell-Klein-Gordon (EMKG) system is described by the action
(with `*' denoting complex conjugate)
\begin{eqnarray}
    \label{action}
    \mathcal{S} =  \int
    d^4 x \sqrt{-g }\bigg[
    \frac{R}{16 \pi G}
    -\frac{1}{4}   F_{ab} F^{ab}
    -\sum_{k=1}^n
       \frac{1}{2} g^{ab}
    \left(
    D_{a}\Psi^{(k)*}  D_{b}\Psi^{(k)} +
    D_{b}\Psi^{(k)*}  D_{a}\Psi^{(k)}
    \right)
    -U (|\Psi| )
    \bigg] ~,
\end{eqnarray}
where $G$ is the gravitational constant, 
$R$ is the Ricci scalar associated with the
spacetime metric $g_{ab}$, 
 $F_{ab} =\partial_a  A_b - \partial_b A_a$ is the Maxwell two-form,
 with $A_a$ the gauge 4-potential.
All  scalar fields are gauged
$w.r.t.$ the same $U(1)$ field,
with the  gauged  derivative
\begin{eqnarray}
    D_{a}\Psi^{(k)} =(\partial_a  + i q  A_a  )\Psi^{(k)}.
\end{eqnarray} 
the potential of the scalar fields, $ U(|\Psi|)>0$,
is  a function of
$|\Psi^{(k)}|^2=\Psi^{*(k)}\Psi^{(k)} $,
and  may
contains interaction terms.

Variation of this action with respect to the metric, gauge potential and scalar 
fields gives the EMKG equations:
\begin{eqnarray}
\label{E-eqs}
&&
 R_{ab}-\frac{1}{2}g_{ab}R= 8 \pi G~
\left[T_{ab}^{(\Psi)} + T_{ab} ^{\rm (EM)} \right],~~
\\
\label{Ms-eqs}
&&
D_{\a}D^{a}\Psi^{(k)}
=\frac{d U}{d\left|\Psi^{(k)}\right|^2} \Psi^{(k)} \ ,  \ \ \ \
\nabla_{a}F^{ b a}=
iq \sum_{k=1}^n\big [\Psi^{(k)*}(D^b \Psi^{(k)})-(D^{b}\Psi^{(k)*}) \Psi^{(k)} )   \big ] \equiv q j^b  ~,
\end{eqnarray}
where  the two components of the energy-momentum tensor are
%
%
\begin{eqnarray}
&&
\nonumber 
T_{ab}^{(\Psi)}=
\sum_{k=1}^n
\left(
 D_{ a}\Psi^{(k)*} 
 D_{ b} \Psi^{(k)}
+ D_{ b}\Psi^{(k)*} 
D_{  a} \Psi^{(k)}
\right)
-g_{ab}  \left[ 
\sum_{k=1}^n
\frac{g^{cd}}{2} 
 \left(
D_{ c} \Psi^{(k)*} D_{ d} \Psi^{(k)}+
D_{ d} \Psi^{(k)*} D_{ c} \Psi^{(k)}
\right)
+U(|\Psi|)
\right],
\\
\label{Tab}
&&
T_{ ab}^{\rm (EM)}
=
F_a^{~c}F_{b c} - \frac{1}{4}g_{ab}F_{cd}F^{cd}~.~~~
\end{eqnarray}
This model is invariant under a $local$ $U(1)$ gauge transformation
$
\Psi^{(k)} \to   e^{-i q \chi(x^\alpha)}\Psi^{(k)},
$
$
A_a\to A_a +\partial_a \chi(x^\alpha)  ,
$
where $\chi(x^\alpha)$ is any real function.
The Maxwell equations in (\ref{Ms-eqs}) define a 4-current $j^a$,
 which is conserved, $\nabla_a j^a=0$.

\section{No (linear) scalar clouds in a dyonic RN background} 
\label{clouds}

Before addressing the full 
EMKG system, it is of interest to
consider a decoupling limit,
solving the 
Klein-Gordon (KG) equation
on the background 
of a solution 
of the Einstein-Maxwell equations,
and thus ignoring the backreaction of the scalars.
The study of this {\it test field limit}
is technically easier and informative about the fully non-linear system.

Restricting to spherical symmetry,
the background corresponds to
 a  RN dyonic  BH  with a line element 
\begin{eqnarray}
\label{RN}
ds^2=-H(r) dt^2+\frac{dr^2}{H(r)}+r^2(d\theta^2+\sin^2\theta d\varphi^2)\ ,~~{\rm with} ~~
H(r)\equiv 1-\frac{2M}{r}+\frac{4\pi G (Q_e^2+Q_m^2)}{r^2}~,~~
\end{eqnarray} 
and a $U(1)$ potential with the nonvanishing components
%
$A_\varphi=Q_m \cos \theta ,$
$A_t =\frac{Q_e}{r} ,$~~
$Q_m$ and $Q_e$
representing the magnetic and electric charge, respectively, while $M$ is the BH's mass.
 Observe that $Q_m$ is not arbitrary, being subject 
to Dirac's quantization condition
 (\ref{Dirac}).
 The metric (\ref{RN})
 possesses an (outer)
 event horizon at
 $r_H=M+\sqrt{M^2-4 \pi G (Q_e^2+Q_m^2)}$.

\subsection{A single scalar field} 
\label{single}
The simplest case here corresponds to a single
scalar field,
$\Psi^{(1)}\equiv \Psi$, 
without self-interaction,
$U (|\Psi|)=\mu^2|\Psi|^2$, where $\mu$ is the field's mass.
Then the KG equation admits separation of variables. 
The expression of
an individual mode of the field is\footnote{In what follows, we shall restrict to the 'plus sign'
in Dirac's condition (\ref{Dirac}).}
\begin{eqnarray}
\label{one-scalar}
\Psi= \psi(r) Y_{N,\ell,m}(\theta,\varphi) 
e^{-i\omega t},
\end{eqnarray}
where $Y_{N,\ell,m}$
are the spherical monopole harmonics,
with
\begin{eqnarray}
\label{Y}
 Y_{N,\ell,m}(\theta,\varphi)=\Theta_{N,\ell,m}(\theta)
 e^{i m \varphi},~~
 {\rm with}~~
 \ell=\frac{N}{2},\frac{N}{2}+1, \dots,~~
 m=-\ell,-\ell+1,\dots, \ell,
\end{eqnarray}
and the expression of $\Theta_{N,\ell,m}(\theta)$ is given, $e.g.$ in~\cite{Wu:1976ge}.  The radial function of a scalar field (\ref{one-scalar})
satisfies the equation 
(where a prime denotes the derivative with respect to $r$):
\begin{eqnarray}
\label{eqU1}
 (r^2 H \psi')'-\mu^2 r^2 \psi
 +\frac{r^2 \psi}{H}\left(w-q \frac{Q_e}{r}\right)^2
 -\lambda \psi=0~, 
\end{eqnarray}
with $\lambda=\ell(\ell+1)-N^2/4>0$ being a separation constant which originates in the angular dependence of $\Psi$.

We are interested in the possible existence 
of bound state solutions
of~(\ref{eqU1}); these
would describe
(linear)
{\it charged scalar clouds}.
For the solutions
to be physically meaningful,
the 
scalar field should be regular on the horizon.
Then, similar arguments to those 
discussed in Section \ref{framework}.
show that the resonance 
condition
(\ref{cond})
still holds, 
with $\Phi=Q_e/r_H$.
A  straightforward computation then casts eq. (\ref{eqU1}) as
\begin{eqnarray}
\label{eqU2}
 (r^2 H \psi \psi')'=
 r^2 \left[
 H\psi'^2+
 \left(
 \frac{\lambda}{r^2}+
 \mu^2 -q^2\frac{Q_e^2}{r_H^2}
+ \frac{r_H^2-4\pi G (Q_e^2+Q_m^2)}{r r_H-4\pi G (Q_e^2+Q_m^2)}  
\right) \psi^2
 \right]~.
\end{eqnarray}
The localization of the field (and the synchronization) impose the condition
$\omega^2 = q^2\frac{Q_e^2}{r_H^2}\leq \mu^2$
while $H'(r_H)\geq 0$
results in 
$r_H^2\geq 4\pi G Q^2$.
Therefore the $r.h.s.$
of the eq. (\ref{eqU2})
is a strictly non-negative quantity outside the horizon.
Then,   integrating 
the relation (\ref{eqU2})
between the horizon and infinity results\footnote{This generalizes for $Q_m \neq 0$
a result in Ref. \cite{Garcia:2021pzd}.
}
in $\psi\equiv0$, using the fact that the $l.h.s.$
part vanishes at infinity,
due to the exponential decay of $\psi(r)$. 

A possible  loophole in the above argument is when
$\mu^2 = q^2\frac{Q_e^2}{r_H^2}=\omega^2$
(maximal frequency),
and
$r_H^2= 4\pi G Q^2$
(extremal RN background), which prevents the exponential decay.  For $Q_m=0$ this
is the case discussed in a more general context in 
Ref. \cite{Degollado:2013eqa}.
However, eq. (\ref{eqU1})
can be solved in this case, with 
the solution
$
 \psi(r)=c_1 (r-r_H)^{\frac{1}{2}(\sqrt{4\lambda+1}-1)}
 +
 c_2 (r-r_H)^{-\frac{1}{2}(\sqrt{4\lambda+1}+1)}~,
$
($c_1,c_2$ being two constants),    
which diverges at the horizon or at infinity, implying no regular bound state solution exists.

It is also interesting to consider the limiting case of
 a massless scalar field with $\omega=0$
 on a purely magnetic RN background.
 The above argument fails in this case, since   the scalar  field may decay asymptotically as $1/r$,
and thus the integral of the $l.h.s$. term in   (\ref{eqU2}) would not vanish. 
However, also in this case eq. (\ref{eqU1})
has an exact solution, with
 $
 \psi(r)=c_1 P_{\frac{1}{2}(\sqrt{4\lambda+1}-1)}
 (\frac{r-M}{\sqrt{M^2-4\pi G Q_m^2}})
 +
 c_2 Q_{\frac{1}{2}(\sqrt{4\lambda+1}-1)}
 (\frac{r-M}{\sqrt{M^2-4\pi G Q_m^2}}),
 $
where $P$ and $Q$ are the Legendre functions,
which again fails to be regular for any non-trivial choice of the 
  constants $c_1,c_2$ 
  and thus
  cannot describe a bound state.

\subsection{ The
multi-scalar case: a specific ansatz} 
\label{multi}

The results of the previous subsection generalize for 
the case of  (suitable) scalar multiplets, $n>1$,
without self-interaction.
 
Let us start by remarking that,
from (\ref{one-scalar}),  (\ref{Y}),
a single gauged scalar field
cannot be spherically symmetric 
 in the presence of a magnetic charge;  additionally, it would carry a
non-vanishing angular momentum density.
Configurations compatible with a spherically symmetric geometry
(and also with a nonzero magnetic charge)
are found, however, in models with
  scalar field multiplets, with a common field mass and a common radial and (harmonic) time  dependence.
The $(\theta,\varphi)$-dependence, on the other hand,  
is specific for each field,
being
chosen such that it vanishes at the level
of the $total$ energy-momentum tensor\footnote{A
  construction in the same spirit exists for 
  (ungauged) boson stars  \cite{Alcubierre:2018ahf}.}.
Then for each
$\Psi^{(k)}$,
the KG equation reduces 
to the same equation for the radial amplitude.

For the case of $N$-monopoles,
the simplest such ansatz\footnote{The ansatz (\ref{gen})
 is not unique; other choices 
 with  $n>N+1$ scalars are possible.}
%
contains $N+1$ scalars, with
\begin{eqnarray}
\label{gen}
\Psi^{(k)}= c_k~ \psi(r)~ \left[\sin\left(\frac{\theta}{2}\right)\right]^{N-k+1} 
\left[\cos\left(\frac{\theta}{2}\right)\right]^{k-1} ~
e^{i( \frac{1}{2}(N-2k+2)\varphi-\omega t)},~~k=1,\dots N+1~,
\end{eqnarray}
where $c_k$ are suitable constants chosen such that
the condition
\begin{eqnarray}
\sum_{k=1}^{N+1}\Psi^{(k)} \Psi^{*(k)}= \psi(r)^2 \ ,
\end{eqnarray}
is satisfied (with $c_k^2=2^N$
and $c_1=c_{N+1}=1$).
The explicit ansatz for $N=1$
is given in eq. (\ref{ansatz-psi})
below; 
for comparison, we include here
the expression for $N=2$:
\begin{eqnarray}
\label{N=2}
\Psi^{(1)}= \psi(r) \left[\sin\left(\frac{\theta}{2}\right)\right]^{2} 
e^{i(\varphi-\omega t)},~~
\Psi^{(2)}=\frac{1}{\sqrt{2}}\psi(r)\sin \theta ~
e^{ -i \omega t},~~
\Psi^{(3)}= \psi(r) \left[\cos\left(\frac{\theta}{2}\right)\right]^{2} 
e^{-i(\varphi + \omega t)}~.
\end{eqnarray}

For any number $N$,
the radial function $\psi(r)$ 
in the Ansatz (\ref{gen}) 
solves the same eq. (\ref{eqU2}), this time,
however, with $\lambda=N$.
This equation
 can also be derived from the reduced action
\begin{eqnarray}
\label{SeffN}
 \mathcal{S}_{red}=\int_{r_H}^\infty~dr
 \left(
H r^2 \phi'^2+\frac{N}{2}\psi^2
-\frac{(\omega- q\frac{Q_e}{r})^2r^2\psi^2}{H}
+r^2 U(\psi)
 \right)~.
\end{eqnarray} 
Taking
$U(\psi)=\mu^2 \psi^2$,
it follows that 
all results 
in subsection (\ref{single}) 
generalize  for the Ansatz (\ref{gen}),
with the absence of 
linear scalar clouds also in this case.

Moreover,
in contrast to the electric case,
one can show additionally that
\textit{no non-linear 
scalar clouds}
exist on a RN background
with a magnetic charge, only.
Taking
$\omega=Q_e=0$
in (\ref{SeffN}),
a Derrick-type scaling technique 
\cite{Derrick:1964ww,Herdeiro:2021teo},
results in the 
virial identity
 \begin{eqnarray}
\label{virial} 
\int_{r_H}^\infty~dr
 \left[
 (r-r_H)^2 \phi'^2+\frac{N}{2}\psi^2 
+r^2\left(3-\frac{2r_H}{r}\right) U(\psi)
 \right]=0~.
\end{eqnarray} 
which, 
for
$U(\psi)$
an arbitrary {\it positive} function
of 
$\sum_k|\Psi^{(k)}|^2$,
implies $\psi=0$, necessarily. 

\section{Spherically symmetric 
dyonic BHs with non-linear scalar hair}
\label{hair}

\subsection{The framework}
\label{framework}

The above no-go results do  not exclude the existence of 
{\it non-linear} clouds on (or backreacting) a \textit{dyonic} BH, which would be necessarily disconnected from the 
dyonic RN BH.
Indeed, 
as shown in~\cite{Hong:2019mcj,Herdeiro:2020xmb,Hong:2020miv},
this is the case for purely electric, spherically 
symmetric configurations,
the solutions being supported by the self-interaction of
the (single) scalar field.
However, for a single scalar field, the 
dependence on the polar angular coordinate $\theta$
 does not factorize for a non-linear potential,
 leading to 
axially symmetric configurations.
This is not the case when considering the
multiplet scalar field ansatz  (\ref{gen}),
provided that the potential $U(|\Psi|)$
is a function of
$\sum_k|\Psi^{(k)}|^2$.
In what follows we shall 
restrict ourselves to the simplest case
with $two$ scalars
and a single monopole, $N=1$.

Employing spherical coordinates with the usual range,
a consistent ansatz reads\footnote{
Ansatz (\ref{ansatz-psi})
has already been 
proposed in~\cite{Brihaye:2023vox}),
 emerging via  a Kaluza-Klein reduction
 of a specific five dimensional scalar doublet.
 Some properties of this scalar ansatz
    (including regularity)
    are discussed in a more
    general context in~\cite{Gervalle:2022npx,Gervalle:2022vxs}.
}
\begin{eqnarray}
     \label{ansatz-psi}
 \Psi^{(1)} = \psi(r) \sin\left(\frac{\theta}{2}\right)~
 e^{i (\frac{\vphi}{2}-\omega t)},~~
  \Psi^{(2)} = \psi(r) \cos\left(\frac{\theta}{2}\right)~
  e^{-i( \frac{\vphi}{2}+\omega t)}~,
\end{eqnarray}
which is compatible with a unit magnetic charge
$Q_m={1}/{(2q)}$,
only.
For the  $U(1)$-field we take a dyonic ansatz 
  \begin{eqnarray}
\label{em}
A=Q_m \cos \theta d\varphi+ V(r) dt,
\end{eqnarray}
$V(r)$ being
the electric potential.
The generic solutions herein  are electrically charged,
since $V=0$ is not a consistent truncation, 
unless $\omega=0$.

As for the metric ansatz, we have found 
convenient to employ isotropic coordinates,
with a  line element containing 
two unknown functions\footnote{The (vacuum) Schwarzschild metric is recovered
for $\mathcal{F}_1=\mathcal{F}_0=0$.}
$\mathcal{F}_0(r)$, $\mathcal{F}_1(r)$:
\begingroup
\footnotesize
\begin{equation}
\label{metric-iso}
 ds^2=-e^{2\mathcal{F}_0(r)}\frac{S_0^2(r)}{S_1^2(r)}dt^2
+  e^{2 \mathcal{F}_0(r)} S_1^4(r)\bigg( dr^2+r^2(d\theta^2+\sin^2\theta d\varphi^2) \bigg),~~{\rm with}~~ 
 S_0(r)=1-\frac{r_H}{r},~~S_1(r)=1+\frac{r_H}{r}.{~~}~~
\end{equation}
\endgroup

We observe that {\it each}
scalar field
carries a non-zero angular momentum density;
however, they cancel at the level of the total energy momentum tensor,
which is compatible with (\ref{metric-iso}).
As such, the equations satisfied by 
the metric and the matter functions can be derived
from the following effective Lagrangian\footnote{In addition, the Einstein equations (\ref{E-eqs}) provide another second order
(constraint) equation, which is not solved directly,
being used to verify the accuracy of the numerical results.}:
\begin{eqnarray}
&&
{\cal L}_{eff}=8 e^{\mathcal{F}_0+\mathcal{F}_1}
\left( \frac{1}{2}r S_1 \mathcal{F}_1'+S_0-1 \right)
\left[ S_0(r)\left( \mathcal{F}_0'+\frac{1}{2} \mathcal{F}_1'\right)-1) +1 \right]
-8\pi G \frac{ e^{\mathcal{F}_0+\mathcal{F}_1} r^2S_0}{S_1^3}
\bigg[
\frac{Q_m^2}{r^4}
{~~~~~~}
\\
&&
\label{LEff}
{~~~~~~~~~~~~~~~~~~}
-\frac{e^{-2\mathcal{F}_0+2\mathcal{F}_1S_1^6}}{S_0^2}V'^2+
2e^{\mathcal{F}_1}S_1^4
\left(
\psi'^2+\frac{\psi^2}{2r^2}
+e^{\mathcal{F}_1}S_1^4 U(\psi)
-\frac{e^{-2\mathcal{F}_0+2\mathcal{F}_1}S_1^6 (\omega-qV)^2}{S_0^2}\psi^2
\right)
\bigg]~.
\nonumber
\end{eqnarray}
  %
 The finite energy requirement  imposes that $\psi \to 0$
 asymptotically
and also that it should possess a mass term, which  localizes the field.

The numerical results  in this work are
found for the
simplest choice of the
scalar potential
in the $Q$-ball literature 
\cite{Coleman:1985ki,Lee:1988ag,Volkov:2002aj,Kleihaus:2005me},
with
\begin{eqnarray}
\label{potential}
U(|\Psi|)=\mu^2\psi^2-\lambda \psi^4+\nu \psi^6~,~~
~~~\psi^2=|\Psi^{(1)}|^2+|\Psi^{(2)}|^2~,
\end{eqnarray}
where $\mu$ is the scalar field mass while $\lambda,\nu$ are
positive parameters controlling the self-interactions of the
scalar field 
(with $\lambda^2 \geq 4\mu^2\nu$, such that $U \geq 0$).
A usual choice in the literature  of the parameters  in 
(\ref{potential})
is $\mu^2=1.1$, $\lambda=2$ and $\nu=1$,
being also employed in 
the numerics herein.
We also briefly note that  no
solutions seem to exist for the choice
$\nu=0$ and $\lambda \leq 0$.

The BH horizon is located at $r=r_H>0$,
where the functions $\mathcal{F}_0,\mathcal{F}_1$
are finite.
Then, requiring finiteness  on the horizon of the energy-momentum
tensor (\ref{Tab})
(or of the current density $j^a$),
implies that the following condition
is satisfied
\begin{eqnarray}
\psi(r_H) [\omega-q V(r_h)] =0 \ .
\end{eqnarray}
If one chooses $\psi(r_H)=0$, this implies that
the derivatives of the scalar field vanish order by order in a power series expansion
close to the horizon, $i.e.$ the scalar field trivializes.
Thus, in order to consider a non-zero scalar field and
finite physical quantities at the horizon, we are forced to consider the second choice,
 $i.e.$  the \textit{resonance condition} (\ref{cond}),
 without
 an explicit contribution of the 
  magnetic charge.

With this choice of the field frequency,
the approximate expression of the solution
near the horizon
reads
\begin{eqnarray}
&&
\label{nh}
\mathcal{F}_0(r)=f_{00}+f_{02}(r-r_H)^2+\dots,~~~
\mathcal{F}_1(r)=f_{10}+f_{12}(r-r_H)^2+\dots,~~
\\
&&
\nonumber
V(r)=\frac{\omega}{q}+v_{2}(r-r_H)^2+\dots,~~~
\psi(r)=\psi_{0}+\psi_{2}(r-r_H)^2+\dots,~
\end{eqnarray}
leaving the free parameters 
$f_{00}$,
$f_{10}$,
$\psi_0$
and
$v_2$,
to fix the coefficients of all higher terms in (\ref{nh}); one finds $e.g.$ 
$\psi_2=\frac{\psi_0}{8r_H^2}+2e^{2f_{10}}\frac{dU(\psi)}{d\psi}|_{\psi=\psi_0}$.
Let us remark that
the
presence
of a magnetic charge clearly excludes regular boson star-like solutions in the limit $r_H \to 0$,
no consistent small-$r$ expansion existing in that case.
This feature can be traced to the presence in 
the field equations of a (Maxwell) term
proportional with 
$F_{\theta \varphi}F^{\theta \varphi}
\sim Q_m^2/r^4 $,
which cannot be regularized in the absence of an horizon.

In the far field region, the scalar field
decays exponentially, 
a RN configuration being approached,
with 
\begin{eqnarray}
\label{inf}
\mathcal{F}_0(r)=\frac{(2r_H-M)}{r}+\dots,~~
\mathcal{F}_1(r)=-\frac{(2r_H-M)}{r}+\dots,~~
V(r)=\frac{Q_e}{r}+\dots,~~
\psi(r)=\frac{c_0}{r} e^{-\sqrt{\mu^2-\omega^2}r}+\dots,
\end{eqnarray}
with $c_0$ a constant,
and $M$, $Q_e$ 
the ADM mass and the electric charge, respectively;
also $\omega\leq \mu$,
which translates into an upper bound for the chemical potential,
$\Phi\leq \mu/q$.

 {\small \hspace*{3.cm}{\it  } }
\begin{figure}[h!]
\hbox to\linewidth{\hss%
	\resizebox{9cm}{7cm}{\includegraphics{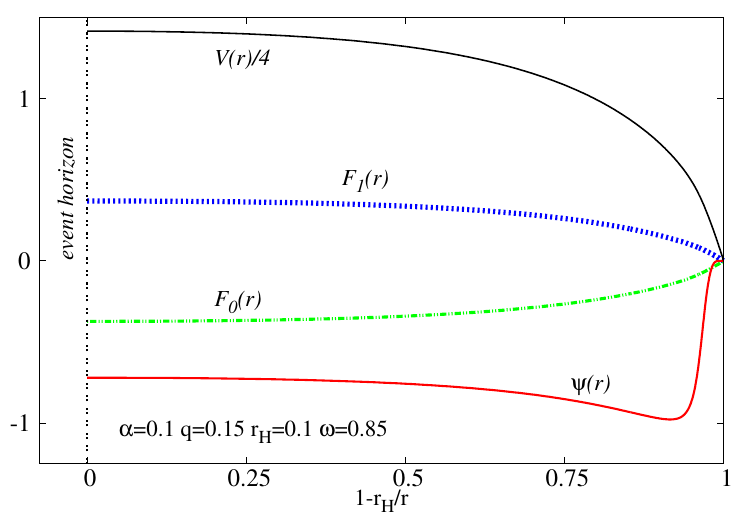}} 
	\resizebox{9cm}{7cm}{\includegraphics{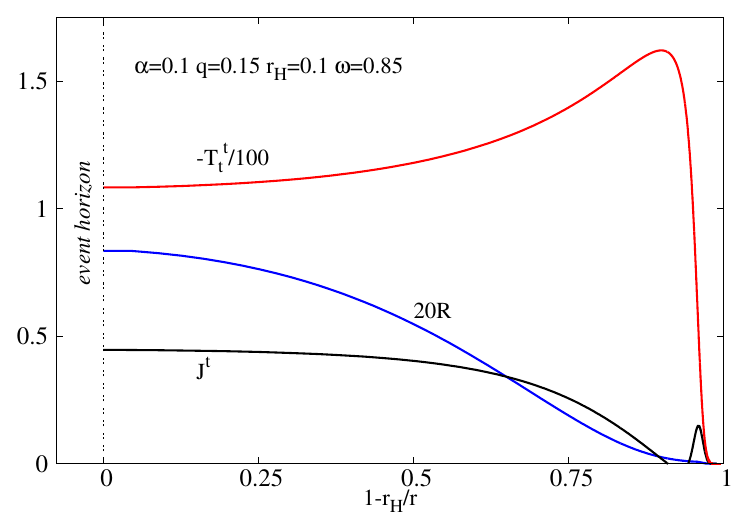}}  
\hss}
\caption{\small  
The profile of a typical dyonic 
BH solution with scalar hair is shown together with 
the corresponding energy density
$\rho=-T_t^t$,
the temporal component of the current
density
$j^t$ and the Ricci scalar $R$.
}
\label{profile}
\end{figure}

Most of the quantities of interest are encoded
in the coefficients which enter the approximate
solutions 
(\ref{nh}),
(\ref{inf}).
The horizon area $A_H$, 
the Hawking temperature $T_H$
and the horizon electric charge $Q_H$ are:
\begin{eqnarray} 
A_H=64 \pi r_H^2 ~e^{2\mathcal{F}_1(r_H)} ,~~
T_H=\frac{1}{64\pi r_H}e^{\mathcal{F}_0(r_H)-\mathcal{F}_1(r_H)},~~
Q_H=\frac{1}{4\pi}\oint_H dS_r F^{rt}=e^{\mathcal{F}_1(r_H)-\mathcal{F}_0(r_H)}r_H^3 v_2~.
\end{eqnarray} 
The electric charge $Q_e$ can also
be expressed as a sum of an
event horizon contribution plus a bulk term,
\begin{eqnarray}
Q_e=Q_H+Q_{(int)},~~{\rm with}~~Q_{(int)}=\int_{r_H}^\infty dr
\frac{ 2U_1}{U_0} e^{-\mathcal{F}_{0}+3 \mathcal{F}_{1}}r^2  (w-q V)\psi^2~.
\end{eqnarray} 
The ratio $Q_H/Q_e$
provides a measure of
 the hairiness of the solutions.

\subsection{Numerical results }

The set of four second order
coupled ordinary differential equations 
for 
$\mathcal{F}_0$, 
$\mathcal{F}_1$,
$V$
and
$\psi$ 
has been solved numerically  subject
to the boundary conditions
$ \psi'(r_H)=
\mathcal{F}'_0(r_H)
=\mathcal{F}'_1(r_H)=0$,
$V(r_H)=w/q$
at the horizon,
while all functions were
imposed to vanish at infinity.
%
In the numerics 
we have made use of a sixth-order finite difference scheme, where the system
of equations is discretized on a grid with a typical size of 
530 points in radial direction.
The emerging system of nonlinear algebraic equations has been solved using the Newton-Raphson scheme.
Calculations have been performed 
by employing a professional solver
\cite{schoen},
with typical errors of order of $10^{-4}$.
Some of the solutions were also constructed by using a standard Runge-Kutta ordinary differential
equation solver. 

The system still possesses a residual scaling symmetry 
\cite{Herdeiro:2020xmb},
which reveals the existence of a
coupling constant $\alpha$
 denoting the strength of the gravitational coupling
 \cite{Kleihaus:2005me}.
As such,
the input parameters of our calculations are  
  $\alpha$, the gauge coupling $q$
($i.e.$ the magnetic charge), the
horizon radius $r_H$ 
and the scalar field frequency $\omega$.
 
The profile of a typical solution 
is shown in Fig.~\ref{profile}.
One can see that the  functions 
${\cal F}_0, {\cal F}_1$ and $V$
are monotonic, while the scalar profile is nodeless and vanishes asymptotically; no sign of a divergent behaviour
is found.

 {\small \hspace*{3.cm}{\it  } }
\begin{figure}[h!]
\hbox to\linewidth{\hss%
	\resizebox{9cm}{7cm}{\includegraphics{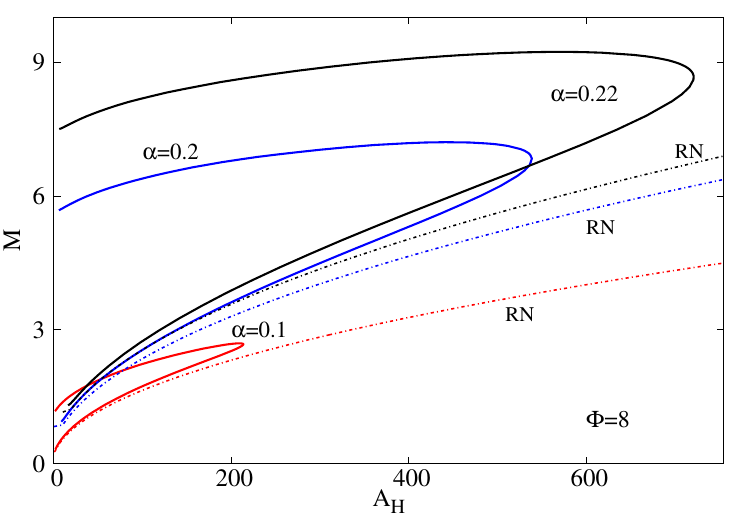}} 
	\resizebox{9cm}{7cm}{\includegraphics{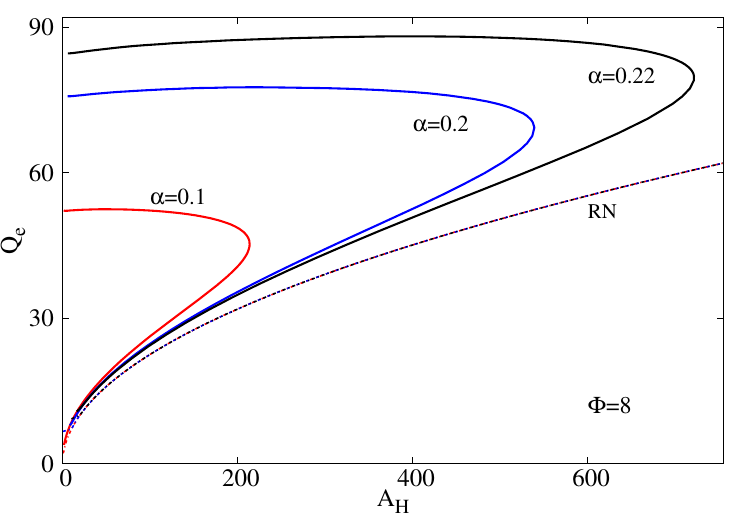}}  
\hss}
\caption{\small 
ADM mass $M$ (left panel) 
and electric charge
$Q_e$ of dyonic
BHs with resonant scalar hair, for a  fixed
chemical potential $\Phi$ and several values of the coupling constant
$\alpha$, $vs.$ the horizon area $A_H$. 
Here and in 
Figs. \ref{gravity2}-\ref{fig2},
the corresponding quantities
for the (bald)  RN  BHs
with the same
input parameters
are also shown
for comparison.
}
\label{gravity1}
\end{figure}
 {\small \hspace*{3.cm}{\it  } }
\begin{figure}[h!]
\hbox to\linewidth{\hss%
	\resizebox{9cm}{7cm}{\includegraphics{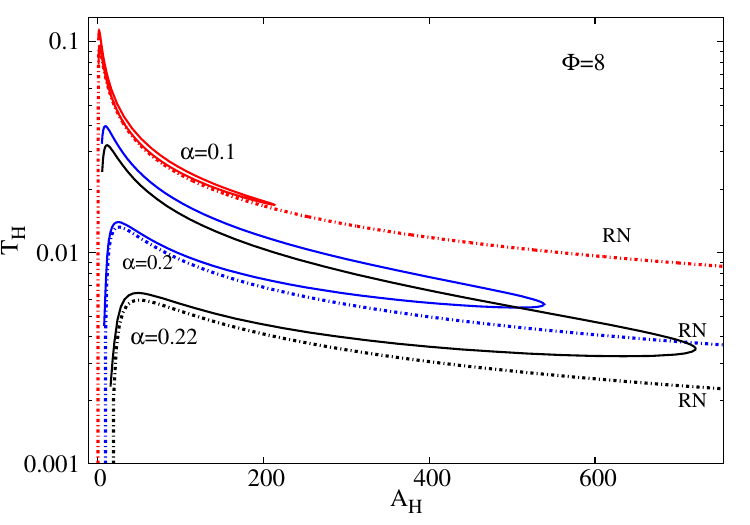}} 
	\resizebox{9cm}{7cm}{\includegraphics{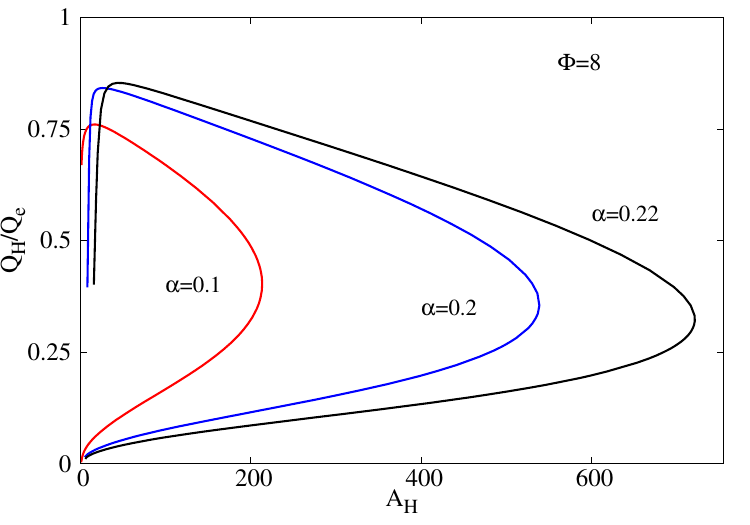}}  
\hss}
\caption{\small 
The Hawking temperature $T_H$ (left panel) 
and the hairiness parameter
$Q_H/Q_e$ (right panel) $vs.$ the horizon area $A_H$, for the same solutions as in Fig.~\ref{gravity1}.
}
\label{gravity2}
\end{figure}

In Fig.~\ref{gravity1} 
we consider the mass and electric charge of hairy BHs
with  the same value of the
electric chemical potential $\Phi=0.8$
and gauge coupling constant $q=0.1$,
in terms of their horizon area $A_H$.
Several values of $\alpha$ have been considered,
with a general pattern emerging.
One notices 
the existence of two branches of solutions,
which join for 
a maximal BH size. 
Along the lower branch, both the mass and the electric charge take values not very different
from those of RN BHs with the same $\Phi$ and $q$.
This is also the case for the  Hawking temperature, see  Fig.~\ref{gravity2} (left panel). 
As $A_H\to A_H^{(\rm max)}$, a secondary branch emerges, with a backbending in $A_H$.  
In Fig.~\ref{gravity2} 
(right panel)
one can see that the hairiness parameter $Q_H/Q_e$ 
never approaches the unit value,
and also does not vanish. 
This is consistent with $i)$ the existence of a mass gap with respect to the bald RN BHs, 
$Q_H/Q_e \neq 1$ (no bifurcation),
and  $ii)$ the absence of a solitonic limit
$Q_H/Q_e \neq 0$.

 {\small \hspace*{3.cm}{\it  } }
\begin{figure}[h!]
\hbox to\linewidth{\hss%
	\resizebox{9cm}{7cm}{\includegraphics{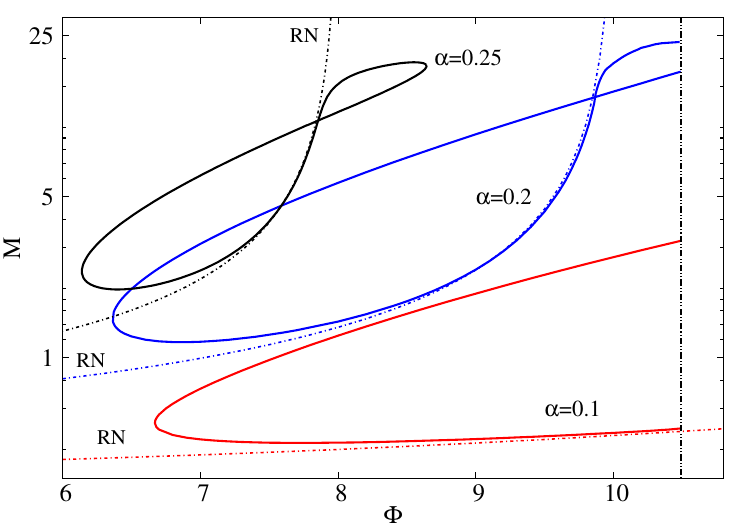}} 
	\resizebox{9cm}{7cm}{\includegraphics{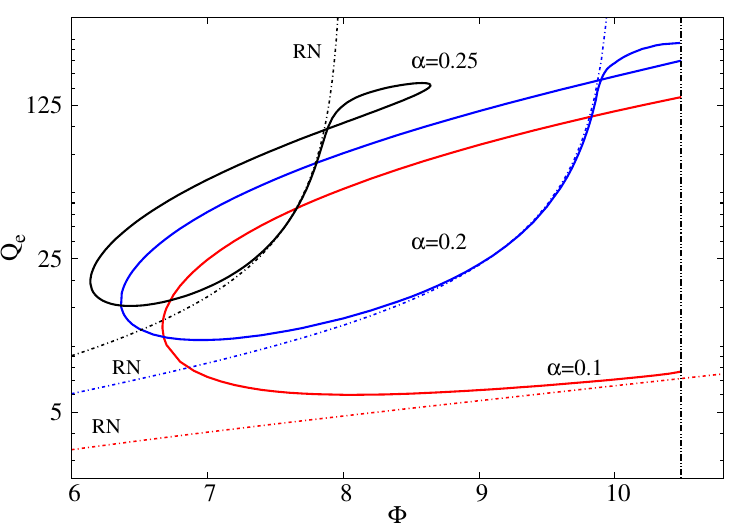}}  
\hss}
\caption{\small  
ADM mass $M$ (left panel) 
and electric charge
$Q_e$ of dyonic BHs with resonant scalar hair, for a  fixed value of the event horizon
radius $r_H=0.1$ and several values of the coupling constant
$\alpha$, $vs.$ the electric 
chemical potential $\Phi$. 
}
\label{fig1}
\end{figure}
 {\small \hspace*{3.cm}{\it  } }
\begin{figure}[t!]
\hbox to\linewidth{\hss%
	\resizebox{9cm}{7cm}{\includegraphics{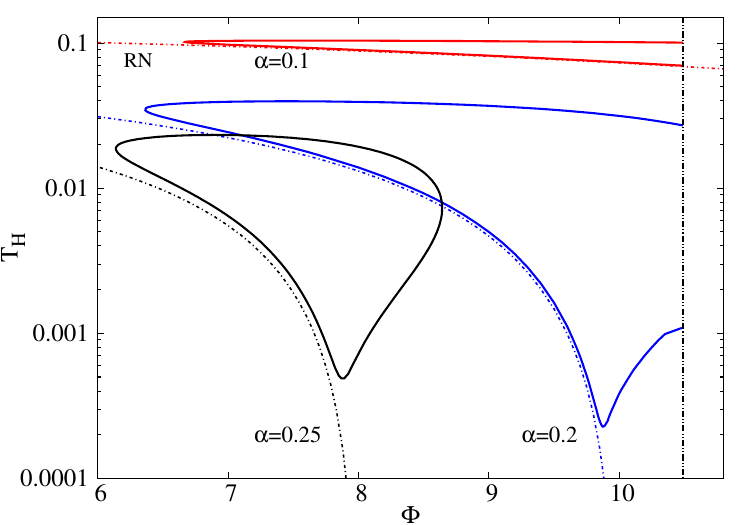}} 
	\resizebox{9cm}{7cm}{\includegraphics{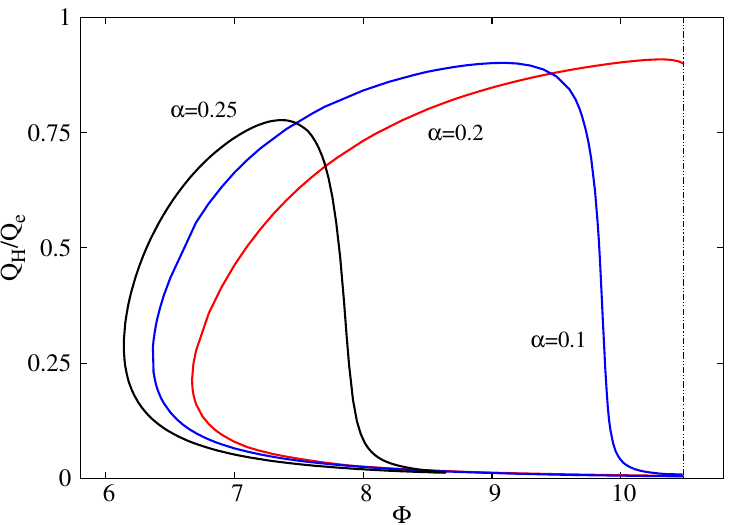}}  
\hss}
\caption{\small 
The Hawking temperature $T_H$ (left panel) 
and the hairiness parameter
$Q_H/Q_e$ (right panel) $vs.$ the electric 
chemical potential $\Phi$, for the same solutions as in Fig.~\ref{fig1}.
}
\label{fig2}
\end{figure}

Although the numerics becomes increasingly challenging
for small $A_H$,
the extrapolated data suggest
that, for all $\alpha$,
both branches start with critical solutions
possessing  nonzero values of
horizon area, 
while their mass and electric charge are finite.
At the same time, the extrapolated Hawking temperature 
appears to vanish,
and thus
we conjecture that the limiting solutions
describe extremal dyonic BHs with scalar hair\footnote{This conjecture is also supported
by the existence of an attractor solution,
with an $AdS_2\times S^2$ geometry and a nonzero
(constant) scalar field.}.
However, the study of these solutions would require a
different metric ansatz, and we hope to report it elsewhere.

A more complicated picture is found when
investigating instead the properties  of solutions
for a varying chemical potential 
at a fixed value
of the horizon radius
(and also a fixed magnetic charge).
The results
in this case are shown in 
Figs.~\ref{fig1}, 
\ref{fig2},
for
several values of 
the coupling constant
$\alpha$
(and  $q=0.1$, $r_H=0.1$).
The picture there reveal a qualitative similarity with the pattern  observed for the
spherically symmetric electrically charged BHs with resonant Q-hair in~\cite{Kunz:2023qfg}.
First, 
the solutions exist within a restricted interval of values of the angular frequency
$\omega \in [\omega_{min},~\omega_{max}]$, 
with the occurrence
of two branches of solutions joining\footnote{Note that $\omega_{min}$ is always non-vanishing,
and thus, as expected, no indication is found for
the existence of
purely magnetic solutions.}
at $\omega_{min}$.
As the gravitational coupling $\alpha$ remains relatively weak,
$\alpha \lesssim 0.21$, the maximal value $\omega_{max}$ corresponds to
the mass of the excitations of the scalar fields, $\omega_{max}=\mu$.
 This picture changes
 for  large enough values of $\alpha$,
 the solutions stopping to exist for some 
 $\omega_{max}<\mu$.
 At that point, a secondary branch emerges, extending 
 backwards in $\Phi$ towards $\omega_{min}>0$,
 the full set of solutions describing a  closed loop 
 in terms
 of $\Phi$.

  In all cases, the lower (fundamental) branch of solutions
follows close enough to the corresponding RN one (with the same  $Q_m$,$\Phi$ and $r_H$).
Also,
on the upper branch the mass of the
 hairy BHs is much higher, and the corresponding
hairiness 
is significantly larger than on the lower 
branch, as shown in Fig.~\ref{fig1}.
As expected, $0<Q_H<Q_e$, 
  the maximal
hairiness being approached 
 for lower branch solutions with maximal $\Phi$.

 \section{Conclusions}
 \label{final}

While the only static, spherically symmetric  
solution
in a model with a gravitating 
 scalar field with a global U(1)-symmetry
 is the (vacuum) Schwarzschild BH, Refs.~\cite{Hong:2019mcj,Herdeiro:2020xmb,Hong:2020miv}
 have shown
 that the situation is different
 for a $U(1)$ gauged scalar field,
 with the existence
 of electrically charged hairy BHs.
The main purpose of this work was to shown 
that these solutions
possess generalizations with an additional magnetic charge,
describing dyonic BHs 
with gauged scalar hair.
Since a 
magnetically charged
configuration with a single gauged scalar field 
cannot be spherically symmetric,
we have considered a more general model with 
a scalar doublet,
which is compatible with 
a unit magnetic charge monopole
and a
spherically symmetric geometry.

An interesting new feature in this case is the absence  of a smooth horizonless solitonic limit of the solutions.
Otherwise  
the dyonic solutions share most of the basic properties
of the BHs with electric charge.
In particular, the resonance condition (\ref{cond})
 between the field frequency and the electric
 chemical potential is still satisfied,
 without a contribution of the magnetic charge.
Furthermore, the  existence of hairy configurations
requires the scalar field to be massive
\textit{and} self-interacting, no such solutions being found for  massive but free scalar fields. 

Most of the existing literature on the subject of
BHs with gauged scalar hair deals with the case of static solutions\footnote{ Kerr-Newman BHs with gauged scalar hair
have been
reported in
Ref. \cite{Delgado:2016jxq}.
However, those solutions possess an electric charge, only (and  no net magnetic charge).}.
 It would be interesting to investigate  the generalizations 
 of the dyonic Kerr-Newman BHs with
gauged scalar hair - see the recent work \cite{Pereniguez:2024fkn}
which discusses the superradiant instability
for a massive
charged scalar field in a magnetic Kerr-Newman background.

\section*{Acknowledgements}
 We thank Nuno Santos for useful comments on the manuscript.
The work of C.H. and E.R.
is supported by the Center for Research and Development in Mathematics and Applications (CIDMA) 
through the Portuguese Foundation for Science and Technology (FCT -- Fundaç\~ao para a Ci\^encia e a Tecnologia)
 through projects: UIDB/04106/2020 (DOI identifier \url{https://doi.org/10.54499/UIDB/04106/2020}); UIDP/04106/2020 (DOI identifier \url{https://doi.org/10.54499/UIDP/04106/2020});  PTDC/FIS-AST/3041/2020 (DOI identifier \url{http://doi.org/10.54499/PTDC/FIS-AST/3041/2020}); CERN/FIS-PAR/0024/2021 (DOI identifier \url{http://doi.org/10.54499/CERN/FIS-PAR/0024/2021}); and 2022.04560.PTDC (DOI identifier \url{https://doi.org/10.54499/2022.04560.PTDC}). Y.S. would like to thank the
Hanse-Wissenschaftskolleg Delmenhorst for support and
hospitality.

\bibliographystyle{unsrt}
\bibliography{biblio}

\end{document}